# Exceptional Points and Asymmetric Mode Switching in Plasmonic Waveguides

Shaolin Ke, Bing Wang, Chengzhi Qin, Hua Long, Kai Wang, and Peixiang Lu

*Abstract*—We investigate the exceptional points (EPs) in a non-Hermitian system composed of a pair of graphene sheets with different losses. There are two surface plasmon polaritons (SPP) modes in the graphene waveguide. By varying the distance between two graphene sheets and chemical potential of graphene, the EPs appear as the eigenvalues, that is, the wave vectors of the two modes coalesce. The cross conversion of eigenmodes and variation of geometric phase can be observed by encircling the EP in the parametric space formed by the geometric parameters and chemical potential of graphene. At the same time, a certain input SPP mode may lead to completely different output. The study paves a way to the development of nanoscale sensitive optical switches and sensors.

## I. INTRODUCTION

Exceptional points are spectral singularities in non-Hermitian systems [1], [2], which possess open boundaries or material loss or gain [3]. The eigenvalues and eigenfunctions of the system coalesce at EPs. In contrast, the degenerate points such as Dirac points in Hermitian systems have same eigenvalues but different eigenfunctions [4], [5]. Apart from bringing about decay or growth of energy, loss and gain can remarkably change the optical features of the system when it operates near EPs. So far many interesting phenomena have been observed in the presence of EPs, such as unidirectional reflection or transmission [6], revival of lasing [7], and loss-induced transparency [8]. The EPs have been also widely investigated in atomic spectra [9], optical and acoustic cavities [10]-[12], photonic crystals [13], multilayer structures [14], and optical waveguides [3].

In the vicinity of EPs, the structures of complex eigenvalues reveal non-trivial topology, which can be studied by encircling EPs in the parametric space [15]. When the system parameters change slowly along a closed loop that enclose an EP, two eigenvalues exchange to each other when the system returns to its initial parameters [16], [17]. Moreover, after experiencing two cycle loops in the same direction, the eigenfunctions acquire an additional phase $\pm\pi$, which is known as geometric phase [18]. Therefore, EP acts as a second-order branch point for eigenvalues, and a fourth-order branch point for the eigenfunctions [16].

The state exchange for slowly encircling EP relies on adiabatic evolution and it has been reported in quasi-static experiments in non-Hermitian system [9], [10]. However, when considering a dynamical encircling, that is, the system parameters changes in time, the evolution of system states don't follow the adiabatic expectation [19]-[21]. Instead, chiral behavior takes place, in the sense that a clockwise loop and an anti-clockwise loop will yield different final states [22], [23].

Beam propagation in a two-mode waveguides with proper modulation is an equivalent platform to explore dynamic encircling. It has been predicted in dielectric waveguide [24] and experimentally realized in microwave metallic waveguide [25]. However, it is not studied in waveguides supporting SPPs. The plasmonic waveguides, which possess deep subwavelength modes, offer new possibilities to engineer light propagation at nanoscale [26]-[28]. The intrinsic losses of SPPs introduce non-Hemitivity into plasmonic waveguide, which will benefits the appearance of EPs. Recent concerns are paid into graphene, a 2D plasmonic material supporting SPPs in infrared and THz ranges [29]. The surface conductivity of graphene can be dynamically tuned via external field or gate voltage [30]. This makes graphene a promising material for tunable optoelectronic devices [31]-[35]. One of the difficulties to explore the dynamical encircling is that two modes associated with EPs should decouple from the other in the system [25]. The number of SPP modes in multilayer graphene waveguide is equal to the number of graphene sheets [36], [37]. Therefore, the double-layer graphene waveguide exactly supports two SPP modes, which will benefit the observation of chiral behavior of dynamical encircling.

In this work, we investigate EPs in a waveguide composed of a pair of graphene sheets with each having different losses. The topological properties of SPPs in the waveguide are demonstrated by adiabatically encircling the EP through smoothly deforming the distance between two graphene sheets and graphene chemical potential. Due to the breakdown of adiabaticity in non-Hermitian system, the asymmetric mode switching is presented, in the sense that encircling EP along a clockwise loop and an anti-clockwise loop leads to different output modes.

## II. LOCATION OF EPS

We start from considering two-level system composed of two neighbored waveguides allowing mode coupling. The amplitude $A(z) = [A_a(z), A_b(z)]^T$ of two waveguides can be described by coupled-mode theory as $idA(z)/dz = HA(z)$ [31], [38], with system Hamiltonian given as

$$H = \begin{bmatrix} \beta_a + i\gamma_a & C \\ C^* & \beta_b + i\gamma_b \end{bmatrix}, \quad (1)$$

where $\beta_a$ and $\beta_b$ denote the real parts of propagation constants for each waveguide, $\gamma_a$ and $\gamma_b$ are the respective loss, and $C$ denotes the coupling constant. The eigenvalues of Eq. (1) take the form

$$\beta_\pm = \frac{\beta_a + \beta_b}{2} + i\frac{\gamma_a + \gamma_b}{2} \pm \sqrt{|C|^2 + [\beta_a - \beta_b + i(\gamma_a - \gamma_b)]^2/4}, \quad (2)$$

An EP arises at the square-root branch point when eigenvalues coalesce with $\beta_+ = \beta_-$. This condition can be fulfilled as $\beta_a = \beta_b$ and $|C| = |\gamma_a - \gamma_b|/2$. Therefore, each waveguide must possess same real part of propagation constants but different losses to obtain an EP.

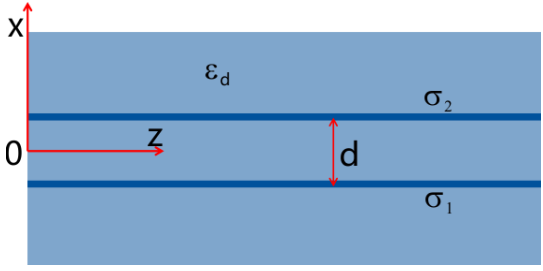

Fig. 1. Schematic diagram of double-layer graphene waveguide. The surface currents of two graphene sheets are denoted as $\sigma_1$ and $\sigma_2$. The chemical potential and relaxation time of lower sheet are fixed at $\mu_1 = 0.15$ eV and $\tau_1 = 0.5$ ps. The relaxation time of upper sheet is chosen as $\tau_2 = 1$ ps, while the chemical potential $\mu_2$ will be tuned. The incident wavelength is at $\lambda_0 = 10$ μm.

The idea can be implemented by using a pair of graphene sheets as shown in Fig. 1, where two graphene sheets are spatially separated with distance $d$. They are embedded in a host dielectric medium with dielectric permittivity $\varepsilon_d$. The graphene sheets possess different surface conductivities labeled as $\sigma_1$ and $\sigma_2$, which relate to wavelength $\lambda$, chemical potential $\mu$, relaxation time $\tau$, and temperature $T$ and can be determined by Kubo formula [35]. Here the room temperature $T = 300$ K is assumed. The incident wavelength is $\lambda = 10$ μm. The intrinsic loss of graphene mainly relates to $\tau$. The relaxation times of two sheets are chosen as $\tau_1 = 0.5$ ps and $\tau_2 = 1$ ps, respectively. Thus the propagation losses of SPPs in the two graphene sheet are different. The chemical potential of lower sheet is fixed at $\mu_1 = 0.15$ eV. To locate EP, two parameters should be tuned. Here, we consider varying chemical potential of upper sheet $\mu_2$ and distance $d$ between two sheets. We only consider TM polarized SPPs and the dispersion relation reads as

$$2 - (\xi_1 + \xi_2)\kappa + \frac{\xi_1\xi_2}{2}\kappa^2[1 - \exp(-2\kappa d)] = 0, \quad (4)$$

where $\kappa = (\beta^2 - \varepsilon_d k_0^2)^{1/2}$ with $\beta$ being the complex SPP propagation constants and $k_0 = 2\pi/\lambda$ denotes the free space wavevector. $\xi_m = \eta_0 \sigma_m/(i\varepsilon_d k_0)$ with $m = 1, 2$ and $\eta_0$ being impedance of air. This double-layer waveguide supports two SPP modes due to the coupling between two sheets. The propagation constants correspond to the system eigenvalues and eigenmodes can be referred as eigenfunctions. Under proper parameters, two complex propagation constants coalesce into one, leading to the appearance of an EP.

Figures 2(a) and 2(b) illustrate the real and imaginary parts of propagation constants as a function of $\mu_2$ when the distance between two sheets is fixed at $d = 127.7$ nm. The results show EP arises at $\mu_2 = 0.15$ eV while Re($\beta$) and Im($\beta$) almost coalesce at the same time. We denote the location of EP as ($\mu_{EP}$, $d_{EP}$) ≈ (0.15 eV, 127.7 nm) in the parametric plane. The SPP propagation constant in single graphene sheet is determined by $\beta_{SPP} = k_0[\varepsilon_d + (2/k_0\xi)^2]^{1/2}$. At an EP, the two sheets almost share equal propagation constants with $\beta_a = \beta_b = 43.4$ μm$^{-1}$, while possess different losses with $\gamma_a = 0.68$ μm$^{-1}$ and $\gamma_b = 0.34$ μm$^{-1}$. Therefore, the complex propagation constant of double-layer graphene waveguide indicated by coupled-mode theory is $\beta_{EP} = (\beta_a + \beta_b)/2 + i(\gamma_a + \gamma_b)/2 = 43.4 + 0.52i$ μm$^{-1}$, approaching the numerical value $43.4 + 0.51i$ μm$^{-1}$ obtained from Eq. (4).

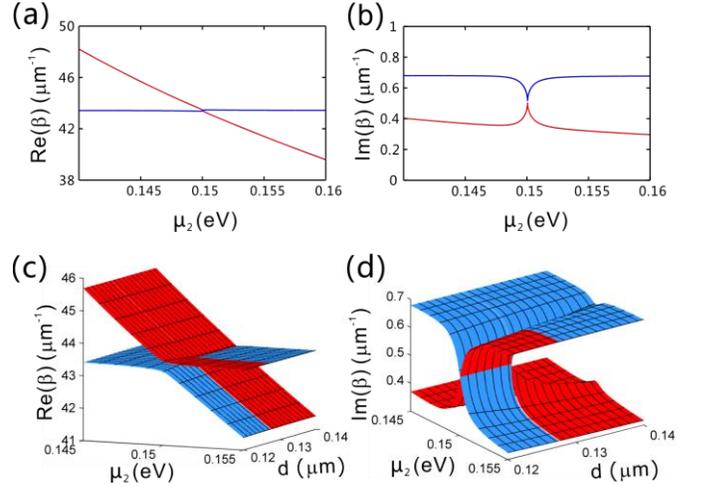

Fig.2. Propagation constants near an EP. (a) and (b) The real and imaginary parts of propagation constants as the chemical potential $\mu_2$ is varying. The period is fixed at $d = 127.7$ nm. The EP locates at (0.15 eV, 127.7 nm) in the parameter space. (c) and (d) Re($\beta$) and Im($\beta$) as a function of $\mu_2$ and $d$.

In Herimitian system, the interacting levels generically avoid crossing (anti-crossing), which corresponds to level repulsion [1]. In contrast, the levels in non-Hermitian systems can experience crossing and avoid crossing in the vicinity of EPs [2]. Figures 2(c) and 2(d) illustrate Re($\beta$) and Im($\beta$) for varying chemical potential $\mu_2$ and distance $d$ in the vicinity of an EP. The EP marks the branch point where the eigen surfaces split. When $d < d_{EP}$, the real part of propagation constants versus $\mu_2$ is anti-crossing, while the imaginary part versus $\mu_2$ is crossing. In contrary, when $d > d_{EP}$, Re($\beta$) shows anti-crossing feature while Im($\beta$) is crossing. This unique topological structure near the EP will allow one to encircle an EP such that two eigenmodes exchange.

### III. TOPOLOGICAL PROPERTY OF EPs

The topological structure of EPs can be studied by encircling them in system parametric space. Two parameters should be turned at the least. The surface current of graphene changes via applying different chemical potentials. This tunable feature will benefit the encircling progress. Here, the chemical potential $\mu_2$ and distance $d$ are varied to investigate the topological property of EPs. The closed loop we choose follows [21], [24], given as a function of parameter $\phi$,

$$d = d_0 + d_1 \sin(\phi/2), \quad \mu_2 = \mu_{EP} + \mu_0 \sin(\phi). \quad (4)$$

When the loop in the parametric space encircles the EP, the condition $d_0 + d_1 > d_{EP}$ should be fulfilled and $\phi$ has to change continuously from 0 to $2\pi$. Here we choose $d_0 = 30$ nm, $d_1 = 80$ nm, and $\mu_0 = 0.02$ eV. The closed loop with a clockwise orientation is shown in Fig. 3(a). The EP locates inside the loop. In Fig. 3(b), the profiles of two eigenmodes $\psi_1$ and $\psi_2$ at the starting point ($\mu_2 = 0.15$ eV and $d = 30$ nm) are illustrated. As the point is far from EP, $\psi_1$ and $\psi_2$ manifest quite different profiles. The main difference between two modes is that $\psi_2$ exhibits very small value at $x = 0$ compared to its maximum of magnetic field, while the value of $\psi_1$ at center of graphene pairs are competitive to its maximum. The corresponding propagation constants for $\psi_1$ and $\psi_2$ are $\beta_1 = 54.1 + 0.45i$ $\mu m^{-1}$ and $\beta_2 = 31.2 + 0.5i$ $\mu m^{-1}$, respectively. The longitudinal SPP propagation distance can be simply written as $L_{SPP} = [2\text{Im}(\beta)]^{-1}$. Then, $L_{SPP}$ for $\psi_1$ is 1.1 $\mu m$ and $\psi_2$ is 1 $\mu m$.

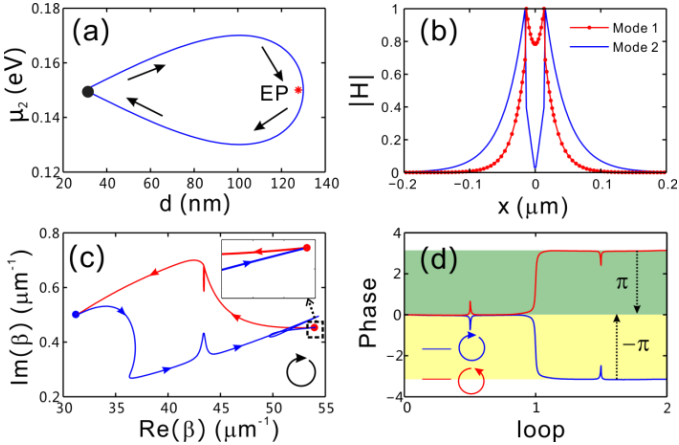

Fig.3. The topological property of EPs. (a) Loop in the parameter space of $\mu_2$ and $d$. The red dot indicates the location of EP. The black dot presents the starting point of loop at (30 nm, 0.15 eV). (b) The field distributions ($|\mathbf{H}|$) at the starting point. (c) The trajectories of propagation constants along the clockwise orientation in the complex plane. The red and blue dots indicate the values of propagation constants at the starting point. (d) Phase evolution of eigenmodes ($H_y$) at $x = 0$. The blue line represents the paramters changes along clockwise loop, while the red one for anti-clockwise loop.

While the system parameters varies clockwise along the loop in the parametric space, as shown in Fig. 3(a), the trajectory of propagation constants are illustrated in Fig. 3(c). Two eigenvalues switch their position at the end of the loop in the complex $\beta$ plane. Thus, two loops with same orientation are necessary for the propagation constants returning to their initial values. This indicates EP serves as a second order branch point for eigenvalues. On the other hand, when EP is not enclosed in the loop, the eigenvalues will return to themselves after one complete loop. During the encircling progress, the corresponding eigenmodes $\psi_1$ and $\psi_2$ exchange their identity after one complete loop as well. Moreover, one of them undergoes a phase change, that is, $\{\psi_1, \psi_2\} \to \{-\psi_1, \psi_2\}$ or $\{\psi_1, -\psi_2\}$. After two cycle loops encircling an EP with same orientation, each mode picks up a phase change, that is, $\{\psi_1, \psi_2\} \to \{-\psi_1, -\psi_2\}$. Therefore, geometric phase appears after encircling an EP twice. In general, eigenmodes possess arbitrary phase obtained from Maxwell equations. We can't directly extract geometric phase from the phase factor of eigenmodes. To explore geometric phase, a continuous phase-plot of eigenfunctions can be used [16], [17]. There are two sets of eigenfunctions in non-Hermitian systems, including the left $\langle\psi_n|$ and right $|\psi_n\rangle$ eigenfunctions. The inner product is defined as $\langle\psi_m|\psi_n\rangle = \int \psi_m^*(x)\psi_n(x)dx$, where $m$ and $n$ represent mode number. The arbitrary phase is removed by requiring the inner product $\text{Im}(\langle\psi_n^*|\psi_n\rangle) = 0$ and $\text{Re}(\langle\psi_n^*|\psi_n\rangle) > 0$. Then geometric phase can be calculated from phase difference between the initial $|\psi_n\rangle_I$ and final states $|\psi_n\rangle_F$ at start and end of parameter loop

$$\left|\Psi_n\right\rangle_F = \exp(i\gamma_B)\left|\Psi_n\right\rangle_I, (5)$$

The wave function can be expressed in the polar form

$$\left|\Psi_n(x)\right\rangle = r(x)e^{i\theta(x)}, (6)$$

The geometric phase is determined by $\gamma_B = \theta_F(x) - \theta_I(x)$, which must be $p\pi$ with an integer $p$. Since $\gamma_B$ is independent of position, the change of phase of an eigenmode can be traced at certain position $x$. The eigenmode here we use is magnetic field ($H_y$) with its initial eigenvalue being $\beta_1 = 54.1 + 0.45i$ $\mu m^{-1}$. The phase evolution at the center of two graphene pairs $x = 0$ is traced as shown in Fig. 3(d). The results show that the loop with different directions will yield different geometrical phase. $\gamma_B$ is $-\pi$ for encircling an EP clockwise after two complete loops, while geometrical phase is $\pi$ for anticlockwise loops. Therefore, the eigenmodes will restore to their initial states after four cycle loops, indicating EP act as a fourth order branch point for eigenmodes.

## IV. ASYMMETRIC MODE SWITCHING

The above state exchange relies on adiabatic evolution. The adiabatic theorem expects that eigenvalues will connect continuously if the system parameters change in small steps [21]. In quantum mechanics, when considering a two-level system with its Hamiltonian $H(t)$ changing in time, the system states evolve according to time-dependent Schrodinger equation (TDSE). In Hermitian systems, the adiabatic solutions converge to the exact solutions of TDSE if system parameters change slowly enough. In contrast, the non-adiabatic terms become significant in non-Hermitian system [20]. As a result, only the least decaying state evolves according to the adiabatic expectation and yields a state change when the system parameters changes in time along a closed loop in the parameter space. At the same time, the other state with higher loss behaves non-adiabatically and returns to itself at the end of the loop [21]. Therefore, the state exchange can't be observed when dynamically encircling an EP. The beam propagation along the waveguide can be regarded as a dynamical evolution as well. Here, the graphene waveguide is modified along the propagation direction according to Eq. (4)

by setting the parameter $\phi = 2\pi z/L$ with $L$ being the propagation length. The beam evolution is now determined by $z$-dependent Hamiltonian corresponding to coupled-mode equation $idA(z)/dz = H(z)A(z)$, a similar form as TDSE.

The distance $d$ and chemical potential $\mu_2$ are varied along $z$ direction, while the surface current $\sigma_1$ remains unchanged and the sheet is fixed at $x = -d_0/2$. The clockwise loop is depicted following Eq. (4), while the anticlockwise loop corresponds to

$$d = d_0 + d_1 \sin(\pi z / L), \mu_2 = \mu_{EP} - \mu_0 \sin(2\pi z / L). \quad (7)$$

The parameters we choose are the same as that used in Fig. 3(a). The slow variation along $z$ direction requires $L|\beta_1-\beta_2| \gg 1$ and we choose $L = 24$ μm.

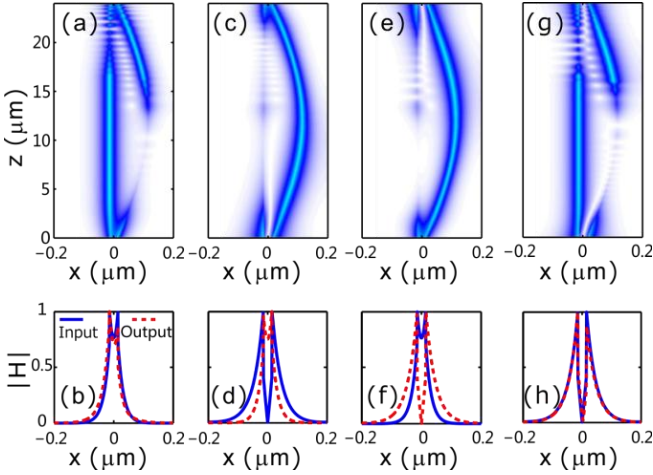

Fig.4. Beam propagation corresponding to dynamical encircling an EP by changing distances $d$ between two sheets and chemical potential $\mu_2$. The parameter in Eq.4 is replaced with $\phi = 2\pi z/L$ and $L = 24$ μm. (a) and (c) is for clockwise loop, using input modes (a) $\psi_1$ and (c) $\psi_2$; (e) and (g) is for anticlockwise loop, excited by modes (e) $\psi_1$ and (g) $\psi_2$. The field is normalized for each $z$ for better visualization. (b), (d), (f), and (h) plot the input (blue line) and output mode (red line) profiles corresponding to (a), (c), (e), and (g), respectively.

The beam propagation is simulated using finite element method (FEM) performed by Comsol Multiphysics. Graphene is modeled as surface currents within the boundary conditions [29], [31]. All the structures are discretized with the mesh less than 1/12 of the SPP wavelength. The input mode at $z = 0$ is the eigenmode $\psi_1$ or $\psi_2$ as shown in Fig. 3(b). The simulation results are illustrated in Fig. 4. The field is normalized for each $z$ for better visualization while the intensity of output mode is actually low. In Figs. 4(a) and 4(c), the beam propagation in the modified waveguides is equivalent to a clockwise loop and input modes are $\psi_1$ and $\psi_2$, respectively. The results show the output mode ($z = L$) is scattering into mode $\psi_1$ for clockwise encircling, regardless of the choice of input mode. On the other hand, the output mode is scattering into mode $\psi_2$ for anti-clockwise encircling as shown in Figs. 4(e) and 4(g). For better visualization, the normalized input and output mode profiles are plotted in Figs. 4(b), 4(d), 4(f), and 4(h), which clearly show the efficient asymmetric mode switching. The asymmetric behavior can be understood as follows. The mode with smaller losses will evolves adiabatically and leads to a mode switching, while the other mode evolves non-adiabatically and returns to itself at the end of loop. The losses refer to the averaging losses over the entire closed loop, given by $1/2\pi \int_0^{2\pi} \text{Im}(\beta)d\phi$ with $\beta$ following the adiabatic expectation as shown in Fig. 3(c). One can see the lower path has smaller losses than the upper one. For a clockwise loop, mode $\psi_2$ evolves along the lower path and we get $\psi_2 \to \psi_1$ and $\psi_1 \to \psi_1$ after a complete loop. In contrast, the evolution of eigenmodes changes their path for an anti-clockwise loop. Then, mode $\psi_1$ follows the lower path and the output modes will be $\psi_2$.

CONCLUSION

In conclusion, we studied the topologic property of asymmetric mode switching by introducing EPs in double-layer graphene waveguides with different losses. By tuning the graphene chemical potential and distance between graphene pairs, the EP appear as propagation constants of two SPP modes coalesce. The geometric phase is observed by adiabatically encircling the EP. We further show the breakdown of adiabaticity when encircling an EP and the asymmetric mode switching is achieved. The approach of dynamical encircling EPs provides an alternative and robust way for mode switching. The study enriches the EP dynamics and paves a way to mode engineering on the deep-subwavelength scale.

ACKNOWLEDGMENT

This work is supported by the 973 Program (No. 2014CB921301), the National Natural Science Foundation of China (No. 11304108), Natural Science Foundation of Hubei Province (2015CFA040), and the Specialized Research Fund for the Doctoral Program of Higher Education of China (No. 20130142120091).

(Corresponding author: B. Wang
e-mail: wangbing@hust.edu.cn).
(Corresponding author: P. Lu
e-mail: lupeixiang@hust.edu.cn).

S. Ke, B. Wang, C. Qin, H. Long, and K. Wang are with School of Physics and Wuhan National Laboratory for Optoelectronics, Huazhong University of Science and Technology, Wuhan 430074, China. (E-mail: wangbing@hust.edu.cn).

P. Lu is with School of Physics and Wuhan National Laboratory for Optoelectronics, Huazhong University of Science and Technology, Wuhan 430074, China, and also with the Laboratory of Optical Information Technology, Wuhan Institute of Technology, Wuhan 430205, China (E-mail: lupeixiang@hust.edu.cn).